\documentclass[pra,aps,twocolumn,floatfix,showpacs,tightenlines,superscriptaddress,amsmath,amssymb]{revtex4}

\usepackage{amssymb,amsbsy,epsfig,color,graphicx}
\usepackage{graphicx}
\usepackage{dcolumn}
\usepackage{bm}

\newcommand{\ket}[1]{|#1\rangle}
\newcommand{\bra}[1]{\langle#1|}
\newcommand{\bk}[2]{\langle#1|#2\rangle}
\newcommand{\eq}[1]{Eq.~(\ref{#1})}

\newcommand{\fig}[1]{Fig.~\ref{#1}}

\begin{document}

\title{Characterization of separability and entanglement in $(2\times{D})$- and $(3\times{D})$-dimensional \\
       systems by single-qubit and single-qutrit unitary transformations}

\author{Salvatore M. Giampaolo}
\affiliation{Dipartimento di Matematica e Informatica, Universit\`{a} degli Studi
di Salerno, Via Ponte don Melillo, I-84084 Fisciano (SA), Italy}
\affiliation{CNR-INFM Coherentia, Napoli, Italy; CNISM Unit\`{a} di Salerno;
and INFN Sezione di Napoli, Gruppo collegato di Salerno, Baronissi
(SA), Italy}

\author{Fabrizio Illuminati}
\thanks{Corresponding author: illuminati@sa.infn.it}
\affiliation{Dipartimento di Matematica e Informatica, Universit\`{a} degli Studi
di Salerno, Via Ponte don Melillo, I-84084 Fisciano (SA), Italy}
\affiliation{CNR-INFM Coherentia, Napoli, Italy; CNISM Unit\`{a} di Salerno;
and INFN Sezione di Napoli, Gruppo collegato di Salerno, Baronissi
(SA), Italy}
\affiliation{ISI Foundation for Scientific Interchange, Villa Gualino, Viale Settimio
Severo 65, I-10133 Torino, Italy}


\begin{abstract}
We investigate the geometric characterization of pure state
bipartite entanglement of $(2\times{D})$- and $(3\times{D})$-dimensional 
composite quantum systems. To this aim, we analyze the
relationship between states and their images under
the action of particular classes of local unitary operations.
We find that invariance of states under the action of single-qubit
and single-qutrit transformations is a necessary and sufficient condition
for separability. We demonstrate that in the $(2\times{D})$-dimensional 
case the von Neumann entropy of entanglement is a
monotonic function of the minimum squared Euclidean distance
between states and their images over the set of single qubit
unitary transformations. Moreover, both in the $(2\times{D})$- and in
the $(3\times{D})$-dimensional cases the minimum squared Euclidean
distance exactly coincides with the linear entropy (and thus as well
with the tangle measure of entanglement in the $(2\times{D})$-dimensional 
case). These results provide a geometric characterization of entanglement 
measures originally established in informational frameworks. Consequences
and applications of the formalism to quantum critical phenomena in
spin systems are discussed.
\end{abstract}

\date{August 3, 2007}

\maketitle

\section{Introduction}

The theory of pure state bipartite entanglement, a trait of quantum
mechanics first discovered more than half a century ago \cite{Einstein}, 
is by now well understood in terms of the entropic content, as first lucidly
pointed out by Schr\"odinger \cite{Schroedinger}, of the reduced
states of a given bipartite quantum system
\cite{VedralPlenioRippinKnight,Vidal,PlenioVirmani,Bengtsson,Horodecki}.
General inseparability criteria are straightforwardly defined for
pure states of composite systems in terms of the factorization of
a given state as a tensor product involving the pure states of
each subsystem \cite{Peres,NielsenChuang}. Besides the direct quantification
via the von Neumann entropy of the reduced states, and the extension to
mixed states via convex roof constructions \cite{Bennett,PlenioVirmani,Horodecki},
it is possible to introduce geometric quantifications of
entanglement in terms of relative entropies \cite{VedralPlenio1,VedralPlenio2,VedralRMP}
or Hilbert space distances defined according to different norms
\cite{Shimony,BarnumLinden,WeiGoldbart}.

An interesting point worth closer investigation is the nature of
the possible relations between entropic and geometric measures of
entanglement and, in parallel, the amount of information on the
global nature of pure states of a composite quantum systems that
can be gained by looking at how states transform under local
operations that, by definition, do not change the content of entanglement (local
unitaries). In fact, besides playing a central role in quantum
computation and quantum information processing, single-subsystem
operations provide global informations on the state of the system.
They find a natural motivation in the operational approach
to the study of physical systems: Looking at the response
to a given action is a basic tool to investigate the nature of
physical properties.

In the present  work we address these questions in the
case of $(2\times{D})$-dimensional and $(3\times{D})$-dimensional
bipartite systems. We investigate the response of pure states to a
particular class of local unitaries 
performed on the two-dimensional subsystem (single-qubit unitary
operations) in the $(2\times{D})$-dimensional case, and on the
three-dimensional subsystem in the $(3\times{D})$-dimensional case
(single-qutrit unitary operations). The specific class of single-qubit 
unitary operations that we consider is the one formed by all
the unitary, Hermitian, and traceless single-qubit
transformations. The analogous class of single-qutrit
unitary operations is the one formed by all the unitary exponentiations
of single-qutrit transformations that are Hermitian, traceless, and
with non degenerate spectrum. This analysis yields that a necessary and sufficient
condition for the separability of pure states is the existence of
at least one element of these classes that leaves the state
unchanged. In fact, this element turns out to be unique up to 
a parity flip. Introducing the standard Euclidean distance between a state
and its image under a transformation, we find that in the $(2\times{D})$-dimensional 
case the entropy of entanglement is a monotonic function of the minimum 
squared distance and that the latter, both in the $(2\times{D})$- and 
$(3\times{D})$-dimensional cases, coincides exactly with the linear entropy 
(or equivalently, in the $(2\times{D}$-dimensional case, with entanglement
monotones such as the squared concurrence and the tangle). 
These results provide a geometric-operational characterization
of entropic measures and entanglement quantifications that were originally
formulated in terms of informational concepts.

The paper is organized as follows: In Section II we introduce the
set of single-qubit unitary operations and analyze the different
transformations that they induce on the pure states of a 
$(2\times{D})$-dimensional bipartite quantum system. We find that pure state
inseparability, or entanglement qualification, can be completely
characterized according to the different ways in which pure states
are transformed by single-qubit local unitaries. Concerning the
quantification of pure state entanglement, in Section III we show
that the von Neumann entropy of entanglement is a single-valued,
monotonic function of the squared Euclidean distance of pure
states from their images, minimized over the entire set of single-qubit 
unitary transformations. In fact, we actually show that this
minimum distance coincides exactly with the linear entropy, or
equivalently the tangle, of the mixed state reductions. This
geometric-operational characterization of entropic measures can be
further connected to the maximization of a measure of the ``local
factorizability'', as discussed in Section IV. In Section V we
generalize the analysis to $(3\times{D})$-dimensional systems. We
again find that a necessary and sufficient condition for pure
state separability is invariance of the states under the action of
single-qutrit unitary operations, and that the minimum squared
Euclidean distance coincides with the linear entropy. We then
establish and discussed the multivalued relation existing between
the linear and the von Neumann entropy. In Section VI we summarize
our findings and briefly remark on the applications of the
formalism, in particular to quantum critical phenomena in spin
systems, and possible further generalizations of the formalism.

\section{Single-qubit unitary operations and separability}

Consider a bipartite quantum system $A|B$ whose associated Hilbert
space is the $(2\times{D})$-dimensional tensor product $\cal{H} =
{\cal{H}}_{A} \otimes {\cal{H}}_{B}$ of the bidimensional space
${\cal{H}}_{A}$ and the $D$-dimensional space ${\cal{H}}_{B}$,
with $D \geq 2$ arbitrary integer. A specific, but important,
realization is given, for instance, by a system  of $N$
distinguishable qubits.
Arbitrarily selecting one of the qubits as subsystem $A$, with
associated bidimensional Hilbert space ${\cal{H}}_{A}$, subsystem
$B$ has associated Hilbert space ${\cal{H}}_{B}$ of dimension
$D=2^{N-1}$. The question we wish to address is whether, and if so
how, one can identify suitably defined unitary operations acting
on ${\cal{H}}$ that, according to the outcome of their actions,
are able to characterize the separability properties of all pure
states $\ket{\Psi} \in \cal{H}$ of the bipartite system $A|B$. In
carrying out this program, we will find that such transformations
allow as well to recast measures of entanglement, such as the
tangle and the von Neumann entropy (single-site entanglement), in
terms of the Euclidean distance between pure states in
${\cal{H}}$.

Proceeding in steps, let us first consider the situation in which
the state $\ket{\Psi}$ is completely factorized with respect to
the bipartition $A|B$:
\begin{equation}
\label{statofattorizzato}
\ket{\Psi} = \ket{\Psi_{F}} = \ket{\Phi_{A}} \otimes \ket{\Xi_{B}} \, ,
\end{equation}
where $\ket{\Phi_{A}} \in {\cal{H}}_{A}$ and $\ket{\Xi_{B}} \in
{\cal{H}}_{B}$. Because subsystem $A$ is a qubit, one can always
write the projection operator $P_{\Phi}^{A}$ on the state
$\ket{\Phi_{A}}$ as:
\begin{equation}
\label{proiettoresuA}
P_{\Phi}^{A} \equiv \ket{\Phi_{A}} \bra{\Phi_{A}} = \frac{1}{2}
\left(\textbf{1}_{A} + \bar{\sigma}_{A} \right) \, ,
\end{equation}
where $\textbf{1}_{A}$ is the identity operator acting on
$\cal{H}_{A}$ and $\bar{\sigma}_{A}$, acting on the same Hilbert
space, is a Hermitian, unitary, and traceless operator such that
$\bar{\sigma}_{A} \ket{\Phi_{A}} = \ket{\Phi_{A}}$, and
$\bar{\sigma}_{A} \ket{\Phi_{A}^{\perp}} = -
\ket{\Phi_{A}^{\perp}}$, where $\ket{\Phi_{A}^{\perp}}$ is the
vector orthogonal to $\ket{\Phi_{A}}$. The operator
$\bar{\sigma}_{A}$ is uniquely defined once the state
$\ket{\Phi_{A}}$ is assigned; equivalently, the latter can always
be defined as the unique right eigenvector of the former with unit
eigenvalue.

Let us now introduce the set of single-qubit unitary operations
$U_{(A|B)}$ ({\em SQUOs}), defined as those unitary
transformations on ${\cal{H}}_{A} \otimes {\cal{H}}_{B}$ that act
separately as the identity $\textbf{1}_{B}$ on ${\cal{H}}_{B}$ and
nontrivially on ${\cal{H}}_{A}$ in the form
\begin{equation}
\label{DefinizioneSQUO}
U_{(A|B)} \equiv {\cal{O}}_{A} \otimes \textbf{1}_{B} \, ,
\end{equation}
where the operator ${\cal{O}}_{A}$ acting on ${\cal{H}}_{A}$ is
such that ${\cal{O}}_{A} = {\cal{O}}_{A}^{\dagger}$,
${\cal{O}}_{A}^{2} = \textbf{1}_{A}$, $\mathrm{Tr}{\cal{O}}_{A} =
0$ (Hermitian, unitary, and traceless). If we choose
${\cal{O}}_{A} = \bar{\sigma}_{A}$, then, from Eqs.
(\ref{proiettoresuA}), (\ref{DefinizioneSQUO}), and
(\ref{statofattorizzato}) we immediately recover that if the state
of the composite system is factorized, then the SQUO
\begin{equation}
\label{SQUOinvariante}
\bar{U}_{(A|B)} = \bar{\sigma}_{A} \otimes \textbf{1}_{B} 
\end{equation}
leaves the state unaltered:
\begin{equation}
\label{Inalterato} \bar{U}_{(A|B)} \ket{\Psi_{F}} =
(\bar{\sigma}_{A} \otimes \textbf{1}_{B}) \ket{\Psi_{F}} =
\ket{\Psi_{F}} \, .
\end{equation}
The SQUO (\ref{SQUOinvariante}) that leaves the separable
state $\ket{\Psi}_{F}$ strictly unchanged will be denoted as the
{\it direct preserving SQUO}, and it is uniquely defined. If we
define the factorized state $\ket{\Psi_{F}^{\perp}}$ orthogonal to
$\ket{\Psi_{F}}$ with respect to qubit $A$:
\begin{equation}
\label{statofattorizzatoOrtogonale} \ket{\Psi_{F}^{\perp}} =
\ket{\Phi_{A}^{\perp}} \otimes \ket{\Xi_{B}} \, ,
\end{equation}
then there exists a unique {\it orthogonal SQUO}
$\bar{U}_{(A|B)}^{\perp}$ that leaves invariant the orthogonal
factorized state $\ket{\Psi_{F}^{\perp}}$:
\begin{equation}
\label{SQUOinvarianteortogonale} \bar{U}_{(A|B)}^{\perp}
\ket{\Psi_{F}^{\perp}} = \ket{\Psi_{F}^{\perp}} \, .
\end{equation}
Applied to the factorized state $\ket{\Psi}_{F}$, the {\it
orthogonal preserving SQUO} (\ref{SQUOinvarianteortogonale})
yields
\begin{equation}
\label{Fattorizzatoribaltato}
\bar{U}_{(A|B)}^{\perp} \ket{\Psi_{F}} = - \ket{\Psi_{F}} \, .
\end{equation}
Therefore, the orthogonal preserving SQUO
(\ref{SQUOinvarianteortogonale}), when applied to the separable
state $\ket{\Psi}_{F}$ produces only an unobservable, absolute
$\pi$ phase shift. Obviously, the reciprocal holds as well:
$\bar{U}_{(A|B)} \ket{\Psi_{F}^{\perp}} = -
\ket{\Psi_{F}^{\perp}}$.

Summarizing, if the state of a composite $(2\times{D})$-dimensional
bipartite quantum system $A|B$ is such that the two-level
subsystem $A$ is decoupled from the $D$-dimensional subsystem $B$,
then there exist two unitary single-qubit operations defined
according to Eqs. (\ref{SQUOinvariante}) and
(\ref{SQUOinvarianteortogonale}) that, respectively, leave the
factorized state unchanged, or modify it by an unobservable global
phase factor $\exp{(i\pi)}$. Viceversa, if the state $\ket{\Psi}$
is entangled, then there is no SQUO that leaves the state
unchanged or modified by an unobservable global phase factor. In
fact, if such a SQUO existed, then $\ket{\Psi}$ would be one of
its two eigenvectors corresponding to one of the two possible
eigenvalues $\pm1$. Hence, it could be put in the form of
\eq{statofattorizzato}, leading to a contradiction with the
initial hypothesis that the state is entangled. Collecting these
results, one has that the existence of a unique SQUO that leaves
the state unchanged is a
necessary and sufficient condition for the factorizability of all
pure states $\ket{\Psi}$ of arbitrary $(2\times{D})$-dimensional
bipartite quantum systems:
\begin{equation}
\ket{\Psi} \, = \, \ket{\Psi_{F}} \; \Longleftrightarrow \;
\exists \; \; U_{(A|B)} \; | \; \; \; U_{(A|B)} \ket{\Psi} \, = \,
\pm \ket{\Psi} \, .
\end{equation}
As a consequence, SQUOs are in principle a useful tool for
the qualification of entanglement in pure states of 
$(2\times{D})$-dimensional bipartite systems by providing a necessary and
sufficient criterion for their separability.

\section{SQUOs, Hilbert space distance, and entanglement}

In this section we demonstrate that the linear entropy, or
equivalently the tangle measure of bipartite
entanglement, for any pure state $\ket{\Psi}$
of a $(2\times{D})$-dimensional system can be rigorously recast in terms of the
Euclidean distance in Hilbert space between the state $\ket{\Psi}$
itself and the state obtained from $\ket{\Psi}$ by transforming it
according to an appropriate SQUO. This result has in turn
interesting consequences on the quantification of entanglement
via observable quantities associated to SQUOs.

Let us begin by briefly recalling some basic notions. Let
$\ket{\Psi}$ be any arbitrary pure state belonging to $\cal{H} =
{\cal{H}}_{A} \otimes {\cal{H}}_{B}$, and $\rho_{A} =
\mathrm{Tr}_{B}(\ket{\Psi}\bra{\Psi})$ the reduced state of
subsystem $A$. The von Neumann entropy (entropy of entanglement)
measuring the bipartite entanglement of state $\ket{\Psi}$ 
is defined as ${\cal{E}}(\ket{\Psi}) =
-\mathrm{Tr}[\rho_{A}\ln\rho_{A}]$. Moreover, for a generic
$l$-dimensional reduced state $\rho_A$, the linear entropy $S_{L}$
is defined as $S_{L} = \frac{l}{l-1}(1 -
\mathrm{Tr}[\rho^{2}_A])$, where the quantity $\mu =
\mathrm{Tr}[\rho^{2}_A]$ measures the {\em purity} of the reduced
state $\rho_A$. In the special case of $l=2$, the linear entropy
can be written as $S_{L} = 4\mathrm{Det}\rho_A$ and the latter
quantity is also known as the {\em tangle}
\cite{Coffman,Osborne2}. 
For the pure states $\{ \ket{\Psi} \}$ of
$(2\times{D})$-dimensional systems, one then has:
\begin{equation}
\label{vonNeumann}
{\cal{E}}(\ket{\Psi}) = - x{\ln}_{2}x - (1 - x){\ln}_{2}(1-x) \, ,
\end{equation}
where $x = (1 + \sqrt{1 - \tau})/2$, and $\tau$ is the tangle 
(or equivalently, the linear entropy) of
the state $\ket{\Psi}$:
\begin{equation}
\label{tangle}
\tau(\ket{\Psi}) = S_{L}(\ket{\Psi}) = 4\mathrm{Det}\rho_{A} \, .
\end{equation}
The von
Neumann entropy \eq{vonNeumann} is thus a simple monotonic
function of the tangle (linear entropy) \eq{tangle}. The latter
varies in the interval $[0,1]$, vanishing for separable states and
reaching unity for maximally entangled states (or equivalently,
maximally mixed reductions with minimal purity $\mu = 1/2$).
It is an entanglement monotone that plays an essential role
in the theory of distributed entanglement and of its ensuing
monogamy property due to the trade off between bipartite and
multipartite nonlocal correlations. In a seminal paper Coffman,
Kundu, and Wootters \cite{Coffman} showed that the tangle is an
upper bound to the sum of all possible pairwise bipartite
entanglements, when measured by the concurrences \cite{HillWootters,Wootters},
for all states of a system of three qubits. Recently,
Osborne and Verstraete \cite{Osborne2} extended the result of
Coffman, Kundu and Wootters by proving that the tangle bounds from
above the sum of all bipartite pairwise entanglement in the
general case of systems of $N$ qubits: The distribution of
bipartite quantum entanglement, as measured by the tangle $\tau$
(or equivalently, by the linear entropy $S_{L}$), satisfies a
tight monogamy inequality \cite{Osborne2}:
\begin{equation}
\label{monogamy}
\tau(\rho_{A_{1}|(A_{2} A_{3} \cdots A_{N})}) \geq
\sum_{i=2}^{N} \tau(\rho_{A_{1} | A_{i}}) \, ,
\end{equation}
where $\tau(\rho_{A_{1} | (A_{2} A_{3} \cdots A_{N})})$ denotes
the bipartite quantum entanglement measured by the tangle across
the bipartition $A_{1} | (A_{2}A_{3} \cdots A_{N})$. As a
consequence of the Schmidt decomposition, the monogamy inequality
\eq{monogamy} holds in general if the system under consideration
is any bipartition $(A|B)$ between a qubit $A$ and a
$D$-dimensional {\em qudit} $B$, with $D \geq 2$.

Let us now consider the Euclidean distance between the state
$\ket{\Psi}$ and the transformed state $\ket{\Psi_{T}}$ obtained
from $\ket{\Psi}$ by acting on it with a SQUO:
$\ket{\Psi_{T}} = U_{(A|B)} \ket{\Psi}$. Denoting such a distance
by $d(\Psi,\Psi_{T})$, we have:
\begin{equation}
\label{distance} d({\ket{\Psi},\ket{\Psi_{T}}}) = \sqrt{1 -
|\bk{\Psi}{\Psi_{T}}|^2} \; .
\end{equation}
Assuming normalized states, the Euclidean distance \eq{distance}
varies in the interval $[0,1]$, vanishing if and only if
$\ket{\Psi_{T}} = \ket{\Psi}$, and reaching unity if and only if
the two states are orthogonal.

After introducing the distance \eq{distance} in $\mathcal{H}$ we
can move on to recast in geometric and quantitative forms the
separability criterion previously established. As we have shown,
subsystem $A$ (qubit $A$) is factorized from the rest of the
system (subsystem $B$) if and only if there exists a unique,
preserving SQUO $\bar{U}_{A|B}$ that leaves the global state
$\ket{\Psi}$ unchanged: $\ket{\Psi}_{T} = \bar{U}_{A|B}\ket{\Psi}
= \ket{\Psi}$. Thus, if subsystem $A$ is factorized, the distance
$d(\ket{\Psi}, \bar{U}_{A|B}\ket{\Psi})$ vanishes. For any
SQUO different from $\bar{U}_{(A|B)} $ and
$\bar{U}_{A|B}^{\perp}$ the transformed state $\ket{\Psi}_{T} =
U_{A|B}\ket{\Psi} \neq \pm \ket{\Psi}$ and $d(\ket{\Psi},
U_{A|B}\ket{\Psi}) > 0$. On the other hand, as soon as qubit $A$
is at least partially entangled with the rest of the system (i.e.
with at least some part of subsystem $B$) we have that any 
SQUO will always change the state $\ket{\Psi}$ and hence, regardless of
the choice of the transformation, $d > 0$. So we have that
necessary and sufficient condition for the factorizability of a
state $\ket{\Psi}$ is that there exists a SQUO for which the distance
between the state and its image under its action vanishes.

Keeping in mind these preliminary results we can go on to quantify
the content of bipartite entanglement for an entangled pure state
$\ket{\Psi}$ by determining the minimum of the distance
\eq{distance} between the entangled state $\ket{\Psi}$ and the set
of its images under all possible SQUOs. Any pure state
$\ket{\Psi}$ of the bipartition $A|B$ can always be put in the
form
\begin{equation}
\label{statogenerico}
\ket{\Psi} = \sum_{n} c_{n,\uparrow} \ket{\uparrow} \ket{n}
+ c_{n,\downarrow} \ket{\downarrow} \ket{n} \; ,
\end{equation}
where $\ket{\uparrow}$ and $\ket{\downarrow}$ stand for the elements 
of an orthonormal basis in the two-dimensional Hilbert space of qubit 
$A$, the set $\{ \ket{n} \}$ forms an orthonormal basis in the
$D$-dimensional Hilbert space of subsystem $B$, and
$\sum_n |c_{n,\uparrow}|^2+|c_{n,\downarrow}|^2=1$. The most
general unitary, Hermitian, and traceless single-qubit operator
${\cal{O}}_{A}$ in the basis formed by $\ket{\uparrow}$ and
$\ket{\downarrow}$ can be cast in the form
\begin{eqnarray}
\label{traceless}
{\cal{O}}_{A} & = & \left(
\begin{array}{cc}
\cos\theta & \sin \theta e^{-i \varphi} \\
\sin \theta e^{i \varphi} & -\cos\theta \\
\end{array}
\right) \\
&& \nonumber \\
& = & \cos{\theta} \sigma_{A}^{z} + \sin{\theta} \cos{\varphi}
\sigma_{A}^{x} + \sin{\theta} \sin{\varphi} \sigma_{A}^{y} \nonumber
\end{eqnarray}
where $\sigma_{A}^{\alpha}$ are the Pauli matrices associated to
qubit $A$, $\theta$ varies in $[0,\pi]$, and $\varphi$ in
$[0,2\pi]$. Thus the distance \eq{distance} associated to any
SQUO is a two-dimensional function $d(\theta,\varphi)$ of the
angular variables $\theta,\varphi$ that parameterize the
single-qubit rotations ${\cal{O}}_{A}$.

Exploiting relations \eq{statogenerico} and \eq{traceless} in
\eq{distance}, we can evaluate the squared Euclidean distance
between the entangled state $\ket{\Psi}$ and its image under any
arbitrary SQUO:
\begin{eqnarray}
\label{squareddistance}
d^{2}(\theta,\varphi) & = & 1 - |\bk{\Psi}{\Psi_{T}}|^{2}
= 1 - |\bra{\Psi}U_{(A|B)}\ket{\Psi}|^{2} \nonumber\\
& = & 1 - \left( M_{z} \cos{\theta} +
M_{x} \sin{\theta} \cos{\varphi} \right.   \\
& & \left. \; \; \; \; \; \;  + M_{y} \sin{\theta} \sin{\varphi}
\right)^{2} \; , \nonumber
\end{eqnarray}
where the expectations $M_{\alpha} = \bra{\Psi}
\sigma_{A}^{\alpha} \ket{\Psi}$ read
\begin{eqnarray}
\label{valorimedi}
 M_{z} & = & \sum_n |c_{n,\uparrow}|^2-|c_{n,\downarrow}|^2 \; , \nonumber \\
 M_{x} & = & \sum_n c_{n,\uparrow} c_{n,\downarrow}^*+c_{n,\downarrow} c_{n,\uparrow}^* \; , \\
 M_{y} & = & -i \sum_n  c_{n,\uparrow} c_{n,\downarrow}^*-c_{n,\downarrow} c_{n,\uparrow}^* \; .
 \nonumber
\end{eqnarray}
Minimizing the distance \eq{distance} (or, equivalently, the
squared distance \eq{squareddistance}) over the entire set of
SQUOs, i.e. over the set $\{\theta,\varphi\}$, we find that the
unique absolute minimum is reached for two different pairs of
values $(\tilde{\theta},\tilde{\varphi})$, corresponding,
respectively, to the extremal SQUO ${U_{(A|B)}}^{extr}$ and
to its orthogonal SQUO ${U_{(A|B)}}^{extr,\perp}$:
\begin{eqnarray}
\label{minimum}
\tilde{\varphi}_1 & = & \arctan{\left( \frac{M_{y}}{M_{x}}\right) } \; , \nonumber \\
\tilde{\varphi}_2 & = & \pi + \arctan{\left( \frac{M_{y}}{M_{x}}\right) } \; , \\
\tilde{\theta}_{1,2} & = & \arctan{\left( \frac{M_{x}\cos{\tilde{\varphi}_{1,2}} +
M_{y}\sin{\tilde{\varphi}_{1,2}}}{M_{z}} \right) } \; . \nonumber
\end{eqnarray}
It is immediate to observe that the two different extremes
correspond to the same absolute minimum for the distance (or
equivalently, the squared distance), because these two quantities
identify two transformed states that coincide but for a global
phase factor. Given the squared Euclidean distance
\eq{squareddistance}, considering its minimum
$\min_{\{\theta,\varphi\}} d^{2}(\ket{\Psi},\ket{\Psi_{T}}) \equiv
d^{2}(\tilde{\theta}_{1,2},\tilde{\varphi}_{1,2})$ yields
\begin{equation}
\label{minimumsquareddistance}
d^{2}(\tilde{\theta}_{1,2},\tilde{\varphi}_{1,2}) = 1 - \left(
M_{x}^{2} + M_{y}^{2} + M_{z}^{2} \right) \; .
\end{equation}
This is exactly the expression of the linear entropy $S_{L}$ (or
equivalently, the tangle $\tau$) \eq{tangle} when cast in
terms of the spin expectation values \cite{AmicoRoscilde}. In
conclusion, for all pure states of $(2\times{D})$-dimensional
quantum systems, one has
\begin{equation}
\label{tangleminimumsquareddistance}
S_{L}(\ket{\Psi}) = \tau(\ket{\Psi}) =
\min_{\{\theta,\varphi\}} d^{2}(\ket{\Psi},\ket{\Psi_{T}}) \; .
\end{equation}
The minimum of the Euclidean squared distance between an entangled
pure state and the set of its images under SQUOs, coincides with
the linear entropy and with the tangle measure of the total
entanglement between qubit $A$ and the rest of the system
\cite{Coffman,Osborne2}. As an obvious corollary, the von Neumann
entropy measure of pure state entanglement is a simple monotonic
function of the minimum squared Euclidean distance
\eq{minimumsquareddistance}. These findings provide a geometric
characterization and interpretation, in terms of single-qubit
unitary transformations and Euclidean distances in Hilbert space,
of entropy and entanglement measures such as the linear entropy,
the tangle, and the von Neumann entropy.

Before ending this section we illustrate an alternative derivation
of the above results that will be useful in the analysis of the 
$(3\times{D})$-dimensional case. In the minimization procedure we have
taken advantage of the fact that SQUOs may be parameterized, see
\eq{traceless}, as functions of the two real parameters $\theta$
and $\varphi$. However, independently of the choice of the
parameters, any SQUO is characterized by the fact that it must
possess two eigenvalues equal to $\pm1$, with associated
eigenvectors
$\ket{+}=\cos(\theta/2)\ket{\uparrow}+e^{i\varphi}\sin(\theta/2)
\ket{\downarrow}$ and
$\ket{-}=\sin(\theta/2)\ket{\uparrow}-e^{-i\varphi}\cos(\theta/2)
\ket{\downarrow}$. Hence, the minimization can be carried out in
in two different but equivalent ways: The one illustrated above,
in which a given basis is fixed and the SQUO is a function of two
real parameters, and a second one in which a given form of the
SQUO is fixed, for instance as
\begin{equation}
\label{traceless1}
{\cal{O}}_{A}= \left(
\begin{array}{cc}
1 & 0 \\
0 & -1 \\
\end{array}
\right) \; ,
\end{equation}
and the orthonormal basis can  be varied as a function of the two
real parameters $\theta$ and $\varphi$. Obviously, by virtue of
the unitary equivalence of representations, the results that are
obtained following the two different procedures coincide. In the
basis of the eigenvectors of the SQUO \eq{traceless1}, the state
$\ket{\Psi}$ can be written as $\ket{\Psi} =
\sum_{n} c_{n,+} \ket{+} \ket{n} + c_{n,-} \ket{-} \ket{n}$. 
Given this expression for the generic state, and taking into account 
the expression of the SQUO in \eq{traceless1} we obtain for the
squared Euclidean distance between the initial and the transformed
state the following expression
\begin{equation}
\label{distance-2-1}
d^{2}(\ket{\Psi},\ket{\Psi_T}) =
1 - (\rho_{11} - \rho_{22})^2 = 4\rho_{11} \rho_{22} \; ,
\end{equation}
where $\rho_{11}=\sum_n|c_{n,+}|^2$ and
$\rho_{22}=\sum_n|c_{n,-}|^2$ are the diagonal elements of the
reduced density matrix $\rho_A$. It is straightforward to observe
that expression \eq{distance-2-1} coincides with the linear
entropy $S_{L}$ (or the tangle $\tau$) if the state $\ket{\Psi}$
is expressed in the basis formed by the eigenvectors of the
reduced density matrix $\rho_A$. Therefore what we are left to
prove is that the minimum of the expression \eq{distance-2-1} over
all possible bases in the two-dimensional Hilbert space is
actually reached in the basis of the eigenvectors of $\rho_A$.
This is easily seen by recalling the trivial fact that the
determinant of a matrix is invariant under a change of basis. Let
us denote by $\tilde{\rho}_{11}, \tilde{\rho}_{22}$ the matrix
elements of $\rho_A$ in the basis of its eigenvectors, and by
$\rho_{\alpha\beta}'$ the matrix elements of $\rho_A$ in a
different basis, arbitrarily chosen. Comparing the expressions of
the determinant of $\rho_A$ in the two bases, one has
\begin{equation}
\label{distance-2-2}
\tilde{\rho}_{11}\tilde{\rho}_{22} = \rho_{11}'\rho_{22}' - |\rho_{12}'|^{2}
\le \rho_{11}'\rho_{22}' \; ,
\end{equation}
and the proof follows.

\section{Linear entropy and tangle as projection operators on pure states}

In the previous section we have showed that the tangle and the
linear entropy coincide with the minimum squared Euclidean
distance between a state and the set of its images under the
action of SQUOs. The key point in the evaluation of the distance
\eq{distance} is the calculation of the real expectation value of
$U_{(A|B)}$ that can be always put in the form
\begin{equation}
\label{forma} \bk{\Psi}{\Psi_T}=\bra{\Psi}U_{(A|B)}\ket{\Psi} =
\mathrm{Tr}(\rho U_{(A|B)}) \; ,
\end{equation}
where $\rho$ is the density matrix (projector) associated to the
pure state $\ket{\Psi}$: $\rho=\ket{\Psi}\bra{\Psi}$. Taking into
account that any SQUO acts as the identity for the degrees of
freedoms that belong to subsystem $B$, we may simply trace out all
these degrees of freedoms to obtain
\begin{equation}
\label{formazza}
\bra{\Psi}U_{(A|B)}\ket{\Psi} = \mathrm{Tr}(\rho_{A}{\cal{O}}_{A}) \; ,
\end{equation}
where $\rho_{A}$ is the reduced density matrix of qubit $A$. The
unitary, Hermitian, traceless operator ${\cal{O}}_{A}$ can always
be expressed as a linear combination of Pauli matrices, with
eigenvalues $\pm 1$ and eigenvectors $\ket{\pm}$. Thus
\begin{equation}
\label{spectral} {\cal{O}}_{A} = \ket{+}\bra{+} - \ket{-}\bra{-}
\; .
\end{equation}
We can then write
\begin{equation}
\label{alternative} \bra{\Psi}U_{(A|B)}\ket{\Psi} =
\mathrm{Tr}(\rho_{A} \ket{+}\bra{+}) - \mathrm{Tr}(\rho_{A}
\ket{-}\bra{-}) \; .
\end{equation}
Since the states $\ket{\pm}$ form a complete basis set for the
subsystem A, the traces $\mathrm{Tr}(\rho_A\ket{+}\bra{+})$ and
$\mathrm{Tr}(\rho_A\ket{-}\bra{-})$ are not independent and
satisfy the relation
\begin{equation}
\label{uffa} \mathrm{Tr}(\rho_{A} \ket{+} \bra{+}) +
\mathrm{Tr}(\rho_{A} \ket{-}\bra{-}) = 1 \; .
\end{equation}
Exploiting \eq{uffa}, relation \eq{alternative} can be
rewritten in the form
\begin{eqnarray}
\label{finale}
\bra{\Psi}U_{(A|B)} \ket{\Psi} & = & 2\mathrm{Tr}(\rho_{A}
\ket{+} \bra{+}) - 1 \nonumber \\
& = & 2\mathrm{Tr}(\rho_{A} \rho_{A}^{p}) - 1 \; ,
\end{eqnarray}
where $\rho_{A}^{p}$ denotes the density matrix associated to a
pure state $\ket{+}$ defined in the qubit $A$. Reminding the
expression of the Euclidean distance \eq{distance}, we have that
\begin{equation}
\label{proiezione}
d^{2}(\ket{\Psi},\ket{\Psi_{T}}) = 1 - \left( 2 \mathrm{Tr}(\rho_{A}
\rho_{A}^{p}) - 1 \right)^{2} \; .
\end{equation}
\eq{proiezione} shows that minimization of the squared distance in
\eq{proiezione} is equivalent to maximizing the ``local
factorizability'' or ``pure state overlap'' $F_{A} \equiv (2
\mathrm{Tr}(\rho_{A} \rho_{A}^{p}) - 1)^{2}$, i.e. the projection
of the reduced density matrix onto a pure state. The local
factorizability varies in the interval $[0,1]$, vanishing for a
maximally entangled state ($\rho_{A}$ maximally mixed) and
attaining unity for a separable state ($\rho_{A}$ pure). Thus, for
an entangled state $\ket{\Psi}$ the nonvanishing minimum
attainable by the squared distance $d^{2}$ is equivalent to the
non-unity maximum attainable by the local factorizability $F_{A}$,
and it is finally straightforward to prove that
\begin{equation}
\label{tanglemaximumoverlap}
S_{L} = \tau = \min_{\{\theta,\varphi\}} d^{2} = 1 - \max_{\{\theta,\varphi\}}F_{A} \; .
\end{equation}
This expression establishes the reciprocal relations existing
between the entanglement of pure states, the minimum squared
Euclidean distance of an entangled state from the entire set of
its images under all possible SQUOs, and the maximum over the set
of all possible SQUOs of the pure-state overlap of single-qubit
reductions.

\section{Single-qutrit unitary operations and entanglement in 
         $(3\times{D})$-dimensional systems}

Let us consider the case of $(3\times{D})$-dimensional composite
quantum systems. Due to the rapidly rising number of parameters
with the local dimension of subsystem $A$ and the involved 
generalizations of the Bloch sphere construction 
\cite{Krauss,Alicki,Byrd,Kimura}, we will follow the strategy of 
minimization outlined in the last part
of Section III by keeping a fixed given expression for the single-qutrit 
transformation and then spanning over all
orthonormal bases in the Hilbert space of the qutrit. We first
need to define the class of single-qutrit unitary transformations
({\em SQUTUOs}). We have seen that the SQUOs have three defining
properties: Unitarity, Hermiticity, and Tracelessness. The last
property obscures the fact that it is equivalent to require that
the spectrum be non degenerate and that the difference between 
any two eigenvalues be a common invariant. In the case of
SQUTUOs, this request must be imposed explicitly (only in the
two-dimensional case this requirement is automatically satisfied
by imposing the condition of vanishing trace) and is fundamental 
in order for SQUTUOs to distinguish between all the different 
components of a state and to weight them on equal footing. This
need, together with the conditions of unitarity and tracelessness
forces the class of SQUTUOs to be formed by non Hermitian
operators. In fact, unitarity implies that for every eigenvalue
$\lambda_k$ ($k=1,2,3$) of a local unitary acting on the
three-dimensional subsystem $A$ one necessarily has
$|\lambda_k|^2=1$ and this relation, together with all the other requirements, 
imposes that the eigenvalues of a given, fixed local unitary be, for instance,
of the form $e^{i \phi}$, $e^{i (2\pi/3 +\phi)}$, and $e^{i
(-2\pi/3 +\phi)}$, where $\phi$ is a global phase factor that can
be set to zero without loss of generality.

We can now proceed to define the form of the associated SQUTUO
acting on a $(3\times{D})$-dimensional space, for which the local part 
of the transformation, acting on the three-dimensional subspace, is
unitary, traceless, and with non degenerate spectrum of
eigenvalues:
\begin{equation}
\label{SQUTUOsdefinition-1}
U_{(A|B)} = \exp \left( i \frac{2 \pi}{3} 
\hat{O}_{A} \otimes \textbf{1}_{B} \right) \; ,
\end{equation}
where $\hat{O}_{A}$ is an hermitian and traceless operator whose
matrix form, in the basis of its eigenstates, reads
\begin{equation}
\label{SQUTUOsdefinition-2}
\hat{O}_{A} = \left(
\begin{array}{ccc}
  1 & 0 & 0 \\
  0 & 0 & 0 \\
  0 & 0 & -1 \\
\end{array}
\right) \; .
\end{equation}
This form coincides, in analogy to \eq{traceless1},
to the matrix form of the operator corresponding to
the $z$ component of the spin-1 subsystem.

Having selected a particular form of the SQUTUOs, we must look at
the action of this operator on a generic state defined in a 
$(3\times{D})$-dimensional Hilbert space. As for the case in which party 
$A$ was a qubit, any pure state $\ket{\Psi}$ of the bipartition $A|B$, with
party $A$ being a qutrit, can always be cast in the form
\begin{equation}
\label{statogenerico3}
\ket{\Psi} = \sum_{n} c_{n,+} \ket{+} \ket{n}
+ c_{n,0} \ket{0} \ket{n} + c_{n,-} \ket{-} \ket{n} \; ,
\end{equation}
where the set $\{ \ket{n} \}$ forms an orthonormal basis in the
$D$-dimensional Hilbert space of subsystem $B$. The set $\{
\ket{+}, \ket{0}, \ket{-} \}$ is formed by the eigenstates of
$\hat{O}_{A}$ associated, respectively, to the eigenvalues 1, 0,
$-1$, and is a basis in the three-dimensional Hilbert space of
qutrit $A$. Finally, $\sum_n
|c_{n,+}|^2+|c_{n,0}|^2+|c_{n,-1}|^2=1$. The action of the SQUTUO
transforms the generic state $\ket{\Psi}$ in the state ${\Psi_T} =
U_{(A|B)} \ket{\Psi}$ that reads
\begin{equation}
\label{statogenerico3-1}
\ket{\Psi_T} = \sum_{n} e^{i\frac{2
\pi}{3}} c_{n,+} \ket{+} \ket{n} + c_{n,0} \ket{0} \ket{n} +
 e^{-i\frac{2\pi}{3}} c_{n,-} \ket{-} \ket{n}.
\end{equation}
Hence
\begin{eqnarray}
\label{statogenerico3-2} \bk{\Psi}{\Psi_T} & = &e^{i\frac{2
\pi}{3}} \sum_{n}
|c_{n,+}|^2 + \sum_{n} |c_{n,0}|^2+ e^{-i\frac{2\pi}{3}} \sum_{n} |c_{n,-}|^{2} \nonumber \\
& = &e^{i\frac{2 \pi}{3}} \rho_{11} +
\rho_{22}+ e^{-i\frac{2\pi}{3}} \rho_{33} ,
\end{eqnarray}
where $\rho_{11} = \sum_{n} |c_{n,+}|^{2}$, $\rho_{22} = \sum_{n}
|c_{n,0}|^{2}$, and $\rho_{33} = \sum_{n} |c_{n,-}|^{2}$ are the
diagonal elements of the reduced density matrix $\rho_{A}$
obtained by tracing over all the degree of freedom of the
$D$-dimensional subsystem $B$. From \eq{statogenerico3-2},after
some elementary algebra, it is straightforward to obtain the
expression of the squared Euclidean distance between $\ket{\Psi}$
and the transformed state $\ket{\Psi_T} = U_{(A|B)} \ket{\Psi}$:
\begin{eqnarray}
\label{distance-3}
d^{2}(\ket{\Psi},\ket{\Psi_T})&=&1-\frac{1}{2}\left[
(\rho_{11}-\rho_{22})^{2} \; + \right .\nonumber \\
& & \left.
(\rho_{11}-\rho_{33})^2+(\rho_{22}-\rho_{33})^2\right] \nonumber \\
&=&\frac{3}{2}\left[1-\left(\rho_{11}^2+\rho_{22}^2+\rho_{33}^2
\right) \right] \; .
\end{eqnarray}
In analogy with the $(2\times{D})$-dimensional case, we are left
with the task of minimizing the expression of the squared
Euclidean distance \eq{distance-3}. This amounts to maximizing the
sum of the squared diagonal elements of the reduced density matrix
$\rho_{A}$ over all possible bases.
\begin{figure}[!t]
\includegraphics[width=6.5cm]{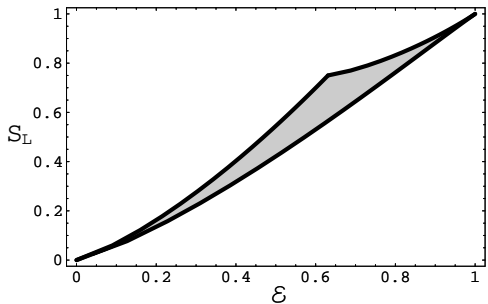}
\caption{\label{linearvsvonneumann} Range of values of the linear
entropy $S_{L}$ at given von Neumann entropy ${\cal{E}}$ for pure
states of $(3\times{D})$-dimensional bipartite quantum systems. The two quantities
coincide at the two extreme cases of separability and of
maximal entanglement. 
The lower boundary line of the interval of 
values for the linear entropy corresponds to the case of two eigenvalues 
of the reduced density matrix $\rho_{A}$ of qutrit $A$ being equal and 
smaller than the remaining one ($\gamma_1 = \gamma_2 < \gamma_3$). 
The left upper boundary line corresponds to 
one vanishing eigenvalue and the remaining two eigenvalues being distinct
($\gamma_1 = 0$, $\gamma_2 \neq \gamma_3$). The right upper boundary 
line corresponds to two coinciding eigenvalues greater than the 
remaining one ($\gamma_2 = \gamma_3 > \gamma_1$). The lower and upper left
boundary lines match at separability; the lower and upper right boundary
lines match at maximal entanglement; the upper left and upper right boundary
lines match when one eigenvalue vanishes and the other two are equal. This point
corresponds to the case in which a two-dimensional subsystem (subspace) of
qutrit $A$ is perfectly entangled with subsystem $B$. All quantities being
plotted are dimensionless.}
\end{figure}
It is easy to show that such maximum is achieved in the basis
formed by the eigenvectors of the reduced density matrix
$\rho_A$. Observing that the trace of $\rho_{A}^{2}$ is invariant 
under changes of basis, we denote by
$\tilde{\rho}_{\alpha,\alpha}$ the matrix elements of
$\rho_A$ in the basis of its eigenvectors 
and by $\rho_{\alpha\beta}'$ the matrix
elements of $\rho_A$ in a different basis, arbitrarily chosen.
Comparing the expressions of $Tr(\rho_A^2)$ in the two bases, one
has
\begin{eqnarray}
\label{distance-3-2}
\tilde{\rho}_{11}^2+\tilde{\rho}_{22}^2+\tilde{\rho}_{33}^2 & = &
\rho_{11}'^2+\rho_{22}'^2+\rho_{33}'^2 \nonumber \\
& + & 2 (|\rho_{12}'|^2+ |\rho_{13}'|^2+
|\rho_{23}'|^2) \; , 
\end{eqnarray}
so that
\begin{eqnarray}
\label{distance-3-4}
\tilde{\rho}_{11}^2+\tilde{\rho}_{22}^2+\tilde{\rho}_{33}^2 & \ge
& \rho_{11}'^2+\rho_{22}'^2+\rho_{33}'^2 \; ,
\end{eqnarray}
and the proof follows. Hence the minimum of the squared Euclidean
distance reads
\begin{equation}
\label{distance-3-5}
\min_{U_{(A|B)}} d^{2} (\ket{\Psi},\ket{\Psi_T}) = \frac{3}{2}
\left[1-\left(\gamma_1^2 + \gamma_2^2 + \gamma_3^2
\right) \right] \; ,
\end{equation}
where the $\gamma_i$ are the eigenvalues of the reduced density
matrix $\rho_{A}$ ($\gamma_i=\tilde{\rho}_{ii}$). 
From the definition of the linear entropy we finally have:
\begin{equation}
\label{minimumdistancequtrit}
S_{L}(\ket{\Psi}) = \min_{U_{(A|B)}} d^{2} (\ket{\Psi},\ket{\Psi_T}) \; .
\end{equation}
Thus the minimum of the squared Euclidean distance between pure
states of $(3\times{D})$-dimensional composite quantum systems and
their images under single-qutrit unitary operations coincides with
their linear entropy, in full correspondence with the result
obtained in Section III in the case of $(2\times{D})$-dimensional
systems. Vanishing of the Euclidean distance, corresponding to the
vanishing the linear entropy or, equivalently, to the reduced
state $\rho_{A}$ of the qutrit being pure, is again a necessary
and sufficient condition for separability. One interesting difference
with the $(2\times{D})$-dimensional case is however that now the linear
entropy is not an entanglement monotone as it is not in one-to-one 
correspondence with the von Neumann entropy (entropy of entanglement). 

Comparison between the two measures, plotted in 
\fig{linearvsvonneumann}, reveals that they coincide at
the lower and upper extrema of separability and maximal entanglement,
while in the intermediate region the linear entropy varies in a 
restricted range at fixed von Neumann entropy. 
The lower boundary line of the interval of 
values for the linear entropy corresponds to the case of two eigenvalues 
of the reduced density matrix $\rho_{A}$ of qutrit $A$ being equal and 
smaller than the remaining one ($\gamma_1 = \gamma_2 < \gamma_3$). 
The left upper boundary line corresponds to 
one vanishing eigenvalue and the remaining two eigenvalues being distinct
($\gamma_1 = 0$, $\gamma_2 \neq \gamma_3$). The right upper boundary 
line corresponds to two eigenvalues equal and greater than the remaining one
($\gamma_2 = \gamma_3 > \gamma_1$). The two lines match at the cusp point 
associated to one vanishing and two coinciding eigenvalues. This situation
corresponds to the case in which a two-dimensional subsystem (subspace) of
qutrit $A$ is perfectly entangled with subsystem $B$.
Qualitatively, this behavior of the linear and von Neumann
entropies for the pure states of $(3\times{D})$-dimensional bipartite 
quantum systems is similar to that of the negativity versus the 
concurrence for the mixed states of two qubits \cite{Verstraete},
and is a particular case of the general comparison between R\'enyi
entropies of different order for fixed $N$-point discrete probability
distributions \cite{Zyczkowski}.

One might think to go further along this line to investigate the effects of
single-ququartet unitary operations on pure states of $(4\times{D})$-dimensional 
bipartite quantum systems, and so on, but, unfortunately,
in these higher-dimensional cases the very definition of the Euclidean
distance between a state and its images is not unique, as it becomes 
dependent on the ordering of the eigenvalues of the local unitary
transformation. Such a problem does not arise if subsystem $A$ consists
of a qubit or a qutrit: In fact, in these cases, the quantities involved
do not depend on the ordering of the eigenvalues.

\section{Consequences, applications, and outlook}

In the present work we have discussed two different questions that in
fact turn out to be related. One concerns the possibility of
characterizing global entanglement properties of composite quantum
systems by suitably defined local unitary transformations (that, by 
definition, do not change the entanglement of a given state). The
other regards the nature of the relationships between entropic and
geometric measures of entanglement. By defining a suitable class
of local unitaries, the single-qubit transformations that are
simultaneously unitary, Hermitian, and traceless, we have found that a
simple necessary and sufficient condition for inseparability stems
from the invariance properties of pure states under such
transformations. This result is straightforwardly generalized to
the $(3\times{D})$-dimensional case. We have then showed that single-subsystem
unitary operations allow for a simple geometrical reformulation of the von
Neumann entropy and other known entropic measures. In particular, both
in the $(2\times{D})$- and $(3\times{D})$-dimensional cases, we have 
demonstrated that the minimum squared Euclidean distance 
between pure states and their images under single-subsystem unitary
transformations coincides exactly with the linear entropy. In turn,
this result implies that in the $(2\times{D})$-dimensional case 
the entropy of entanglement is a unique, single-valued, and monotonic
function of the minimal Euclidean distance. 

Besides the
conceptual interest, these findings can be of use in applications,
for instance in measuring and characterizing the separability properties 
and entanglement content of the ground state $\ket{G}$ of generic, translationally-invariant 
models of interacting qubits (two-level systems or spin $1/2$) or qutrits (spin $1$), 
both at and away from criticality \cite{NostroPreprint}. In particular, such a characterization
can be obtained operationally by introducing an observable, the {\it energy of excitation}
$\Delta E$ above the ground state, that is associated to single-qubit or single-qutrit unitary 
transformations and is defined as \cite{NostroPreprint}: $\Delta E = \bra{G}O^\dagger{\cal{H}}O\ket{G} 
- \bra{G} {\cal{H}} \ket{G}$, where $O$ is a shorthand for the single-spin unitary operation
between an arbitrarily selected spin (subsystem $A$) and the rest of the system (subsystem
$B$), and ${\cal{H}}$ denotes the Hamiltonian of the system. This excitation energy is positive
defined, vanishes if and only if the ground state is factorized, and is a monotonic function of
the linear entropy (or equivalently of the tangle in the $(2\times{D})$-dimensional case). 
Aside from the characterization of ground-state properties at the approach of
quantum critical points from a quantum informatic standpoint \cite{FazioReview},
and the relations to single-site entanglement \cite{Johannesson}, we expect that 
the method of local unitary operations may be of particular importance as a general
technique to ascertain the existence and the location of factorizability
points, i.e values of the Hamiltonian parameters for which the ground state (and, in general,
any pure state) factorizes and becomes classical in models of interacting many-body quantum 
systems \cite{VerrucchiReview}. 

Concerning the extension of the methods discussed in this work to higher-dimensional 
situations and to mixed states, the present strategy will certainly need to be modified 
and integrated by developing more sophisticated tools, including the use of multiple 
operations to yield properly defined distances between states and generalizations to 
mixed-states fidelities and metrics \cite{Inpreparation}. However, some important 
simplifications occur if one restricts the analysis to the case of multimode pure Gaussian 
states of infinite-dimensional systems, for which the framework introduced in the present 
work can be successfully applied with only minor modifications \cite{Gaussian}.

\acknowledgments

We acknowledge financial support from MIUR under PRIN National Project 2005, from INFN, CNR-INFM
Coherentia, CNISM-CNR, and ISI Foundation.

\end{document}